\documentclass[aps,prx,preprint,superscriptaddress,nofootinbib,longbibliography]{revtex4-2}
\usepackage{amsmath,amssymb,bm}
\usepackage{graphicx}
\usepackage{xcolor}
\usepackage[colorlinks=true,linkcolor=blue,citecolor=blue,urlcolor=blue]{hyperref}

\newcommand{\Top}{\mathcal{T}}
\newcommand{\dgamma}{\Delta\gamma}
\newcommand{\QSGWhat}{QS$G\hat{W}$}

\begin{document}

\title{Current selects the helicity of a chiral phonon}

\author{Swagata Acharya}
\email{swagata.acharya@nlr.gov}
\affiliation{National Laboratory of the Rockies, Golden, Colorado, USA}
\author{Jerome Jackson}
\affiliation{Scientific Computing Department, STFC Daresbury Laboratory, Warrington WA4 4AD, United Kingdom}
\author{Mark van Schilfgaarde}
\affiliation{National Laboratory of the Rockies, Golden, Colorado, USA}

\begin{abstract}
The emergence of spin-polarized electrons from nonmagnetic chiral molecules and crystals is called Chirality-induced
spin selectivity (CISS).  A chiral phonon of
definite wavevector $q$ can strongly enhance it, but near equilibrium the two
helicities $+q$ and $-q$ are
equally populated: they are time-reversal partners and their contributions
cancel.  We show that in a biased chiral conductor the current causes the
phonon of one helicity to damp less than the other.  The damping asymmetry is
odd in the current, odd under $q\rightarrow-q$, invariant under a mirror that
reverses the structural handedness, and equal, mode for mode, to the
current-induced nonconservative Berry force on the phonon coordinate.  Closing
the phonon kinetics with a rate equation turns the asymmetry into a net lattice
helicity: at any finite bias one helicity is preferentially populated, with no
threshold, growing linearly with the current as $V\rightarrow0$; above a
threshold that helicity becomes a self-sustained coherent travelling wave.  We
demonstrate this
effect in elemental tellurium, treated as an open quantum system (electronic
Hamiltonian, phonon bath, and electron--phonon coupling) within a
quasi-\textit{ab initio} framework.  The electronic Hamiltonian is supplied by
a vertex-corrected quasiparticle self-consistent $GW$ (\QSGWhat{}) potential
with spin--orbit coupling; nuclear displacements are modeled with a
machine-learned interatomic potential, which yields chiral phonon modes at
finite $\pm q_z$ on the $\Gamma$--A line. The perturbation to the
electronic Hamiltonian from a nuclear displacement is approximated by a
frozen-phonon deformation potential.  In tellurium the damping asymmetry grows linearly with the current, the resulting lattice helicity persists from low temperature to room temperature, and the coherent regime appears below about 20~K at 0.2~V and 40~K at 0.4~V.  No optical pump is needed: a bias alone drives the current and selects the
phonon helicity; reversing the current reverses the
selection.
\end{abstract}

\maketitle

\section{Introduction}\label{sec:intro}
Electrons transmitted through a chiral molecule or crystal emerge spin
polarized, even though neither the incident electrons nor the material carries a net
magnetization.  This is called chirality-induced spin selectivity
(CISS).\cite{Ray1999:Science,Gohler2011,Xie2011,Naaman2012:JPCL,Naaman2019:NatRevChem,Evers2022,Bloom2024,vanSchilfgaarde2026:Newton}
First observed in simple molecules, it has since been demonstrated
across self-assembled monolayers, hybrid perovskites, and layered
crystals.\cite{Michaeli2017,Lu2019,Qian2022,Dong2025} It now serves as a
potential framework for spintronics
without ferromagnets and enantioselective enhancement of catalytic and
electrochemical reactions.  Its microscopic origin, however, remains an open question.
Particularly perplexing is the observation that
bare spin--orbit coupling is far too small to produce the observed polarizations on
its own, especially in light-element chiral materials.
Often the polarization's dependence
on temperature and bias is used as a mechanistic probe.\cite{Evers2022,Bloom2024,Klein2023,vanSchilfgaarde2026:Newton}

\begin{figure*}[t]
\includegraphics[width=\textwidth]{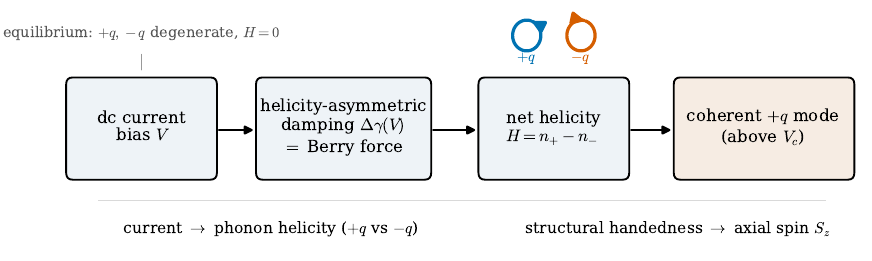}
\caption{Selection of a chiral-phonon helicity by a dc current.
At equilibrium the $+q$ and $-q$ modes are degenerate time-reversal partners and
the net helicity vanishes.  A bias drives the electrons into a steady state that
damps the two helicities unequally; the asymmetry $\dgamma(V)$ is the same
object as a nonconservative Berry force on the phonon coordinate.  A phonon
rate equation converts $\dgamma$ into a net helicity $H=n_{+}-n_{-}$ at any
finite bias, and, above a threshold $V_c$, into a self-sustained coherent
travelling mode of definite $q$.  Structural handedness, through spin--orbit
coupling in the electronic Hamiltonian $H_0$, sets the axial spin $S_z$ that
correlates to the selected mode.}
\label{fig:schematic}
\end{figure*}


Every CISS observable (spin polarization, magnetoresistance, 
spin-to-charge conversion) is odd under time reversal and vanishes in thermal
equilibrium.  A steady signal
therefore requires the system to be driven out of equilibrium, and the size and
sign of the response are controlled by whatever does the driving.  Vibrations
have been proposed as one such driver, since a lattice mode can produce an
effective, spin-dependent exchange splitting through the spin--orbit
coupling.\cite{Fransson2020}  In Ref.~\cite{AcharyaUnified2026} we
organize competing CISS proposals under a single nonequilibrium selection rule
and compare them on one Hamiltonian.  That study covers
more than vibrations: it also maps orbital-to-spin conversion, length and
temperature trends, and the inverse (spin-to-charge) effect. Among the
vibrational treatments considered there (coherent and incoherent drives, and
quantum baths with and without memory), a \emph{coherent chiral phonon} gives
by far the largest, most robust polarization.  A coherent chiral phonon is a
travelling lattice wave of definite
wavevector $q$ whose ionic motion circulates and breaks time reversal on its
own.
Chiral phonons of this kind are now measured directly by circularly polarized
Raman scattering, resonant inelastic X-ray scattering, and ultrafast
probes,\cite{Zhu2018,ZhangNiu2015,Ishito2023,Ueda2023,Tauchert2022,Choi2024}
and they carry a well-defined angular
momentum.\cite{ZhangNiu2014,Juraschek2019}
Related magneto-chiral terahertz dynamics in photoexcited tellurium have
also been reported;\cite{Huang2026:NatPhys} we return to that connection
elsewhere.

A coherent chiral phonon, however, has a definite wavevector $q$, and for this
effect to be manifest, something must cause its population to dominate over
its mirror partner.  In thermal equilibrium the $+q$ and $-q$
modes of a nonmagnetic crystal are degenerate time-reversal partners and
equally occupied, so the lattice supplies \emph{no} net helicity and cannot,
by itself, drive CISS.  Something must select one helicity over the other.
 In the small number of experiments performed under circularly polarized or
terahertz illumination the light itself can do
this,\cite{Zhu2018,Ueda2023,Choi2024,Nova2017} but the overwhelming majority of
CISS measurements are ordinary dc transport experiments with no optical pump at
all.  A current can also induce phonon angular momentum: analytical expressions have been
derived for chiral metals with spin--orbit coupling (the phonon Edelstein effect)
and for supercurrents in superconductors with spin--orbit
coupling.\cite{Yokoyama2025,Yokoyama2026}  How a current selects a particular helicity in these experiments requires an
explanation, which we present here.

We show that the selection is supplied by the
transport current itself.  A dc bias places the electrons in a
nonequilibrium steady state that damps the two helicities unequally.  The
imbalance in damping rate $\dgamma=\gamma_{+q}-\gamma_{-q}$ is odd in the current, odd under
$q\rightarrow-q$, and identical to the
nonconservative, ``current-induced'' Berry force that drives vibrational
instabilities in current-carrying
nanoconductors.\cite{Dundas2009,Todorov2010,Lu2010,Bode2011,Lu2012}  Closing
the phonon kinetics with a rate equation converts $\dgamma$ into a 
lattice helicity in two regimes (Fig.~\ref{fig:schematic}).  At small bias a
net helicity $H=n_{+}-n_{-}$ appears with no threshold, growing linearly in
the current as $V\rightarrow0$, which follows directly from the odd, continuous vanishing
of $\dgamma(V)$ at $V=0$.  This channel exists only
through the electron-phonon coupling, and vanishes in its absence.  It is present at any finite bias, but above a threshold the preferred helicity's net damping
turns negative and it self-amplifies into a coherent travelling wave.
This is the coherent chiral phonon we previously identified as the strongest
vibrational route to CISS.  Ref.~\cite{AcharyaUnified2026} showed that a coherent, self-sustained regime
can produce an anomalously large signal.  In this work, 
the origin of the coherent wave is derived rather than assumed.
We first develop the theory for the model Hamiltonian of Ref.~\cite{AcharyaUnified2026} to establish the effect,
and also show that the current-induced asymmetry emerges in a
quasi-\textit{ab initio} study of elemental tellurium, with
electronic structure, phonons, and their mutual coupling computed from first
principles.  The effect is specific to chiral systems with finite-$q$
chiral phonons; ordinary achiral conductors show none of it. In all these calculations,
Onsager reciprocity is preserved at low bias.

\section{Model}\label{sec:model}
We use a helical tight-binding wire of the Guo--Sun
form~\cite{GuoSun2012,GuoSunPNAS2014}: the minimal ingredients needed for a
chiral conductor.  Sites sit on a helix, each carrying a spin-$1/2$ orbital.
The static Hamiltonian $H_0$ has three physical ingredients: a uniform on-site
energy $\varepsilon_n$, structureless by itself; a nearest-neighbor hopping
$t$; and a chiral spin--orbit coupling of strength $\lambda_{\mathrm{soc}}$,
built from the local torsion of the backbone,
\begin{align*}
H_{0} &= \sum_{n}\varepsilon_{n}\,c_{n}^{\dagger}c_{n}
-t\sum_{n}\bigl(c_{n+1}^{\dagger}c_{n}+\mathrm{h.c.}\bigr) + H_{\mathrm{soc}},\\
H_{\mathrm{soc}} &= i\lambda_{\mathrm{soc}}\sum_{n}c_{n+2}^{\dagger}\,
(\bm{\sigma}\cdot\bm{\nu}_{n})\,c_{n} + \mathrm{h.c.},\quad
\bm{\nu}_{n}=\frac{\bm{a}_{n}\times\bm{a}_{n+1}}{|\bm{a}_{n}\times\bm{a}_{n+1}|},
\end{align*}
with $\bm{a}_n$ the bond vectors and $\bm{\sigma}$ the vector of Pauli
matrices.  Only $H_{\mathrm{soc}}$ ties the electron's spin to the sense of the
helix; it is the only place chirality enters the electronic Hamiltonian.  Two
wide-band leads, held at chemical potentials $\mu_{L,R}=\mu\pm eV/2$, drive a
steady current through the wire.

A phonon of wavevector $q$ modulates the bonds with a travelling phase
$e^{iqn}$ and couples to the electrons through a matrix $A_q$, giving the full
electron--phonon Hamiltonian
\begin{equation}
\mathcal{H} = H_{0} + \hbar\omega_{0}\bigl(b_{q}^{\dagger}b_{q}+b_{-q}^{\dagger}b_{-q}\bigr)
+\!\!\sum_{Q=\pm q}\!\bigl(c^{\dagger}A_{Q}c\,b_{Q}+\mathrm{h.c.}\bigr).
\label{eq:Hfull}
\end{equation}
Here $b_{Q}^{\dagger}$ creates one quantum of the harmonic phonon mode at
wavevector $Q=\pm q$ and bare frequency $\omega_{0}$.  Like $H_0$, the vertex
$A_q$ can have up to three channels: a Holstein term (on-site energy
modulation), a hopping modulation, and a modulation of the chiral
spin--orbit bond, the only one of the three built from the same chirality
vector $\bm{\nu}_n$ as $H_0$.  The calculations below keep the hopping and
chiral spin--orbit channels active and switch the Holstein channel off; the two active channels are comparable in size and interfere, so
neither alone stands in for $A_q$.

The key property of $A_q$ is that it is not Hermitian.  Because the
travelling phase propagates in one direction along the bonds,
$A_{-q}=A_q^{\dagger}\neq A_q$: the vertex for the $-q$ helicity is the
Hermitian conjugate of, not equal to, the vertex for $+q$.  This lets the
electron bath tell the two helicities apart, once driven.  The full Hamiltonian of Eq.~\eqref{eq:Hfull},
vertex and phonon bath included, is Hermitian by construction: every term
appears together with its own h.c.\ partner.  $H_0$ alone, and $H_0$ plus the
phonon coupling with no leads attached at all, are both Hermitian and
time-reversal symmetric at every bias.

For
Hermitian Hamiltonians, Eq.~\eqref{eq:trs} given below must hold.  As a numerical
check we find it does so to machine precision in every calculation we ran.  The
only non-Hermitian object anywhere in this work is the ordinary lead broadening
$-\tfrac{i}{2}(\Gamma_L+\Gamma_R)$: real particle exchange with two reservoirs,
external to the Hamiltonian.  The current enters through a differential in the
leads' chemical potentials, and the effect emerges from the open boundaries under
bias.

The electron-induced damping of mode $q$ is read from the electron--phonon
self-energy evaluated on the driven Green's functions.  Formally, $\Pi_q$ is
the standard one-loop (Migdal) self-energy of the phonon propagator: an
electron loop dressed by two insertions of $A_q$ (see Supplemental materials).  Because the internal
propagators are those of the driven, nonequilibrium steady state, $\Pi_q$ is
purely electronic in origin; the harmonic term in Eq.~\eqref{eq:Hfull}
carries no phonon--phonon interaction of its own.  The intrinsic
(phonon--phonon) lattice damping is therefore a separate, phenomenological
ingredient, $\gamma_{\mathrm{lat}}$, added only in the rate-equation closure
below.  With the greater and lesser phonon self-energies $\Pi^{\gtrless}_q$,
\begin{equation}
\gamma^e_q=\mathrm{Re}\!\left\{i\bigl[\Pi^{>}_q(\omega_0)-\Pi^{<}_q(\omega_0)\bigr]\right\},
\label{eq:gamma}
\end{equation}
with $\gamma^e_q>0$ damping and $\gamma^e_q<0$ antidamping.  We always
write $\gamma^e_q$, not the bare $\gamma_q$, to flag that this is the
\emph{electronic} contribution to a mode's linewidth: the same quantity as
$\gamma^e_\pm$ in the rate equation below, and distinct from
$\gamma_{\mathrm{lat}}$.  Equation~\eqref{eq:gamma} is simply the Migdal
phonon linewidth generalized to a biased steady state.  Its two terms are the
golden-rule absorption and emission rates,
$\Gamma^{\mathrm{abs}}_q\equiv\mathrm{Re}\{i\Pi^{>}_q(\omega_0)\}$ and
$\Gamma^{\mathrm{em}}_q\equiv\mathrm{Re}\{i\Pi^{<}_q(\omega_0)\}$, so that
$\gamma^e_q=\Gamma^{\mathrm{abs}}_q-\Gamma^{\mathrm{em}}_q$.  At equilibrium
detailed balance forces $\Gamma^{\mathrm{abs}}_q>\Gamma^{\mathrm{em}}_q$ for
every mode.  A bias skews the occupation of the electron bath feeding these
golden-rule rates; for the helicity whose vertex is preferentially aligned
with the current-carrying electrons, $\Gamma^{\mathrm{em}}_q$ can grow past
$\Gamma^{\mathrm{abs}}_q$, flipping the sign of $\gamma^e_q$.  The inversion
lives in the electron bath's occupation, not in an
inverted phonon population.

The symmetry content of this needs no calculation.  Time reversal $\Top$
reverses the helicity, $\Top:q\rightarrow-q$, and maps the equilibrium
electronic state onto itself.  At $V=0$, therefore,
\begin{equation}
\gamma_{+q}(0)=\gamma_{-q}(0),
\label{eq:trs}
\end{equation}
an exact degeneracy we confirm numerically to machine precision.  A bias
breaks $\Top$ in the electron bath: reversing $V$ is equivalent to a time
reversal, so the splitting is odd in $V$; relabelling $+q\leftrightarrow-q$ at
fixed bath is also a time reversal, so the splitting is odd in $q$ too.  The
physical selection signal is therefore the bias-antisymmetric part
\begin{equation}
\dgamma(V)\equiv\gamma_{+q}(V)-\gamma_{-q}(V)=-\dgamma(-V),
\label{eq:Dgamma}
\end{equation}
which vanishes identically in equilibrium and grows with the current.



\begin{figure*}[t]
\includegraphics[width=\textwidth]{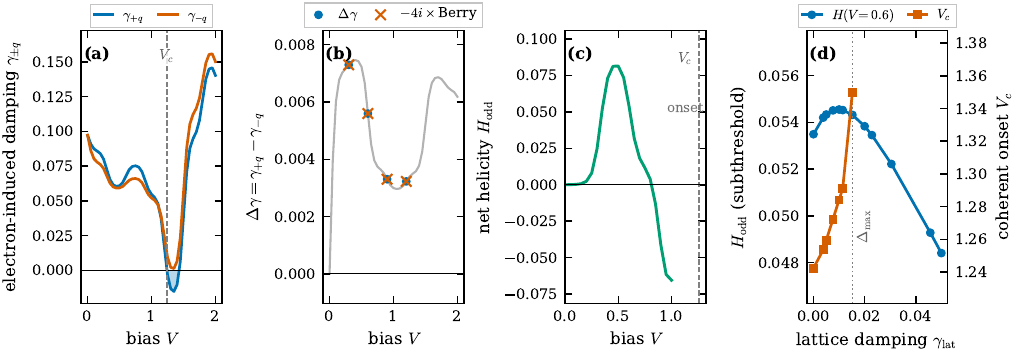}
\caption{Helical-wire model.
(a)~Electron-induced damping of the two helicities versus bias.  They are equal
at $V=0$ by time-reversal symmetry and split under bias; the soft helicity is
driven toward antidamping and crosses zero at $V_c$ (shaded region:
$\min\gamma<0$).
(b)~Helicity asymmetry $\dgamma(V)$ from the standing-wave/circular construction
(grey line), shown for $V\ge0$.  Circles are $\dgamma$ at selected biases;
crosses are the independently computed nonconservative Berry force in the
standing-wave basis ($-4i\times$Berry).  Circles and crosses fall on the grey
line to numerical precision [Eq.~\eqref{eq:identity}].  $\dgamma$ is odd under
$V\to-V$, so the negative-bias half is the mirror image (not plotted);
panel~(a) uses the traveling-wave vertex of the rate-equation closure.
(c)~Net helicity $H_{\mathrm{odd}}$ from the phonon rate equation in the
subthreshold window.  A finite helicity appears with no threshold; near
$V=0$ it grows linearly with the current, then softens (and can change sign)
as $V$ approaches $V_c$, where the coherent mode takes over.
(d)~Phase boundary versus intrinsic lattice damping $\gamma_{\mathrm{lat}}$:
the subthreshold helicity $H_{\mathrm{odd}}$ (left axis) is finite for every
$\gamma_{\mathrm{lat}}$, while a hard coherent onset $V_c$ (right axis) exists
only for $\gamma_{\mathrm{lat}}$ below the maximum electronic antidamping
$\Delta_{\max}$.}
\label{fig:model}
\end{figure*}

\section{Results}\label{sec:results}

\subsection*{Helicity-asymmetric damping is the Berry force}\label{sec:berry}

Figure~\ref{fig:model}(a) shows the two dampings as the bias is swept.  They
coincide at $V=0$, split under bias, and the softer helicity (the one whose
motion is aided by the current) is driven toward antidamping; its
$\gamma^e_q$ crosses zero at a bias we call $V_c$, defined here in the
idealized limit $\gamma_{\mathrm{lat}}=0$.  The rate-equation closure below generalizes $V_c$ to
the bias where the \emph{total} damping, $\gamma^e_q+\gamma_{\mathrm{lat}}$,
first crosses zero: a finite lattice damping raises $V_c$, and removes the
instability altogether once $\gamma_{\mathrm{lat}}$ exceeds the maximum
electronic antidamping $\Delta_{\max}$.  The difference $\dgamma(V)$ constructed from the standing-wave/circular
vertices [Fig.~\ref{fig:model}(b)] is odd in $V$, as required by
Eq.~\eqref{eq:Dgamma}, and rises to a magnitude comparable to the equilibrium
damping itself.  (Panel~(a) uses the traveling-wave vertex of the rate-equation
closure; both constructions are odd in $V$ and vanish at equilibrium.)

We now identify $\dgamma$ with a quantity from a different literature.
In a real (standing-wave) basis the electronic friction felt by the phonon
coordinates is a matrix; a current makes it non-symmetric, and its
antisymmetric part is a nonconservative force that does net work around a cycle
of the ionic coordinates.  This ``current-induced'' or Berry force is the origin
of current-driven vibrational runaway in atomic
wires.\cite{Dundas2009,Todorov2010,Lu2010,Bode2011,Lu2012}  Transforming the
two degenerate standing modes into circular (helicity) combinations, the
antisymmetric friction becomes exactly the difference of the two circular
dampings.  We evaluate the Berry force independently in the standing-wave basis
and find
\begin{equation}
\dgamma=-4i\times(\text{Berry force}),
\label{eq:identity}
\end{equation}
holding to the full numerical precision at every bias we checked
[symbols and crosses in Fig.~\ref{fig:model}(b) are indistinguishable].
Both sides of Eq.~\eqref{eq:identity} are two ways of evaluating the
same electronic friction matrix $\mathrm{Im}\,\Pi$, in two different bases of
the same two-dimensional phonon coordinate (circular for $\dgamma$, standing-wave
for the Berry force); Eq.~\eqref{eq:identity} is therefore an identity between two
representations of one object.
The three descriptions in circulation (helicity-asymmetric phonon damping,
the nonconservative Berry force, and current-induced phonon angular-momentum
pumping) are one object computed three ways.  Nonconservative current-induced
forces are familiar in nanoconductors with more than one vibrational
coordinate;\cite{Dundas2009,Todorov2010,Lu2010,Bode2011,Lu2012} what is special
here is that, in a chiral system, those coordinates are the two circular
polarizations of a phonon, so the force becomes a helicity selector.

Six independent checks fix the physical content of $\dgamma$.  It vanishes at
$V=0$ [Eq.~\eqref{eq:trs}]; it is nonzero and grows under bias; it is odd in
$V$ and odd in $q$; it is \emph{unchanged} when the helix is mirrored at fixed
laboratory $q$, so the helicity that the current selects is locked to the
current and not to the structural handedness; and it is identically zero for an
achiral (straight) wire with ordinary Rashba coupling, which carries a current
but has no chirality to select.  The last two points are the crux of the
experimental phenomenology.  Handedness does enter CISS, but through a
different channel: it sets the axial spin $S_z$ that the selected phonon
imprints, and $S_z$ flips under the same mirror that leaves $\dgamma$
invariant.  The observed dependence of CISS on both bias polarity and molecular
handedness is the product of these two independent sign locks
(Fig.~\ref{fig:schematic}).

\subsection*{From damping to a lattice helicity}\label{sec:rate}

An asymmetry in the damping rates is not yet a population imbalance.  To turn
$\dgamma$ into a physical helicity we close the phonon kinetics.  Each mode
exchanges energy with the electrons at the rates contained in
Eq.~\eqref{eq:gamma} and, in addition, relaxes toward its thermal value
$n_B(\omega_0)$ through the intrinsic anharmonic (lattice) damping
$\gamma_{\mathrm{lat}}$ that any real crystal has.  The occupation of each
helicity then obeys
\begin{equation}
\frac{dn_{\pm}}{dt}
=\Gamma^{\mathrm{em}}_{\pm}
-\gamma^e_\pm\,n_{\pm}
-\gamma_{\mathrm{lat}}\bigl(n_{\pm}-n_B\bigr),
\label{eq:rate}
\end{equation}
where $\Gamma^{\mathrm{em}}_{\pm}$ is the electronic emission rate into the mode
(the lesser self-energy) and $\gamma^e_\pm$ is the net electronic damping of
Eq.~\eqref{eq:gamma}.  In the stable regime, where
$\gamma^e_\pm+\gamma_{\mathrm{lat}}>0$, the steady occupations are
\begin{equation}
n_{\pm}
=\frac{\Gamma^{\mathrm{em}}_{\pm}+\gamma_{\mathrm{lat}}\,n_B}
{\gamma^e_\pm+\gamma_{\mathrm{lat}}}.
\label{eq:nss}
\end{equation}
At $V=0$ the fluctuation--dissipation relation ties emission to damping,
$\Gamma^{\mathrm{em}}_{\pm}=n_B\,\gamma^e_\pm$, and Eq.~\eqref{eq:nss} collapses
to $n_{+}=n_{-}=n_B$ for \emph{any} $\gamma_{\mathrm{lat}}$: no bias, no
helicity.  Under bias the numerator and denominator split between the two
helicities and a net lattice helicity
\begin{equation}
H=n_{+}-n_{-}
\label{eq:H}
\end{equation}
develops.  A finite open wire carries a small bias-even background from its
leads; since the physical selection is odd in the current we report the
projection $H_{\mathrm{odd}}(V)=\tfrac12[H(V)-H(-V)]$, which isolates it.

Two regimes emerge, and they are physically distinct.

\emph{(i) Subthreshold: linear, threshold-free selection.}  For any bias
below $V_c$, $H_{\mathrm{odd}}$ is nonzero and, at small bias, linear in the
current [Fig.~\ref{fig:model}(c)].  There is no threshold: an infinitesimal
current already tilts the balance toward one helicity, because $\dgamma$
itself is already nonzero at any finite $V$ (above), and it is only the
existence of a coherent, self-sustained mode, not the underlying selection,
that requires crossing $V_c$.  We characterize it by a linear
susceptibility $\chi_{\mathrm{PE}}=dH_{\mathrm{odd}}/dV|_{V\to0}$, which is
finite and of order $2\times10^{-3}$ per unit bias in the model.

\emph{(ii) Coherent onset.}  As $V\rightarrow V_c$ the net damping of the soft
helicity, $\gamma^e_++\gamma_{\mathrm{lat}}$, approaches zero and the linear
occupation~\eqref{eq:nss} diverges.  Beyond $V_c$ that mode is antidamped and its
amplitude grows until anharmonicity limits it.  The soft helicity then forms a large-amplitude,
definite-$q$ travelling wave generated by the current and not an assumed input, while the competing
helicity keeps its ordinary occupation.  The coherent regime is the strong-drive limit of the
subthreshold current-induced selection.

The boundary between the two regimes is controlled by a single scale, the
maximum electronic antidamping $\Delta_{\max}=\max_V[-\min(\gamma_{+q},
\gamma_{-q})]$ [Fig.~\ref{fig:model}(d)].  A hard coherent onset requires the
lattice to be soft enough, $\gamma_{\mathrm{lat}}<\Delta_{\max}$; when the
lattice is stiffer than that, the mode never goes unstable and there is no
$V_c$.  The subthreshold helicity, in contrast, is robust: for every
$\gamma_{\mathrm{lat}}$ we tried, $H_{\mathrm{odd}}$ at fixed subthreshold bias
is nonzero.  In a chiral wire the current always chooses a helicity; only the
conversion of that choice into a self-sustained coherent wave is conditional on
the material's anharmonicity.


\begin{figure*}[t]
\includegraphics[width=\textwidth]{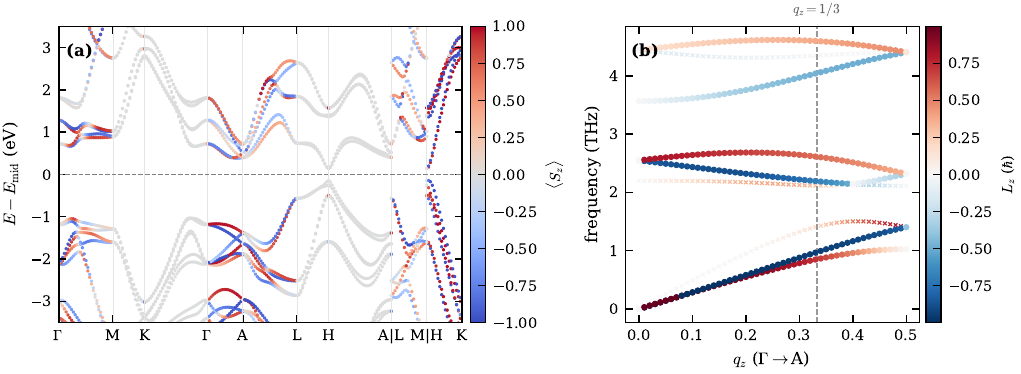}
\caption{\textit{Ab initio} ingredients for elemental tellurium.
(a)~\QSGWhat{}$+$SOC band structure of right-handed tellurium, colored by
the spin expectation value $\langle S_z\rangle$; grey marks $\langle
S_z\rangle\approx0$.  Computed gap $\approx285$~meV.
(b)~Chiral phonon dispersion along $\Gamma$--A, colored by the on-site
angular momentum $l_z$; filled circles mark the six branches with
$|l_z|\gtrsim0.3$ at $q_z=1/3$ (dashed line) used in Fig.~\ref{fig:te};
crosses mark the three that are not.  Near $q_z\approx0.4$ two mid-optical
branches cross; an apparent $l_z$ drop on a frequency-sorted index is a branch
exchange under that sorting (Supplementary Information).}
\label{fig:abinitio}
\end{figure*}
\subsection*{Ab-initio validation of the model framework: Tellurium from first principles}\label{sec:te}

The model isolates the mechanism with the minimal set of ingredients.
To test whether the same current-induced selection survives in a real material
with a real band structure, real spin--orbit coupling, and real chiral phonons,
we repeat the construction for elemental tellurium.  The electronic
structure, the phonons, and their mutual coupling are computed from first
principles.  The transport
setup is specified by the chemical potential, the transverse momenta, the number of
cells in the wire, the temperature and the lead coupling in the wide-band limit (Methods).  At the H point and $V=0.2$~eV, $\dgamma_{\mathrm{odd}}$ changes by a factor of about 2.6 when the lead coupling is varied from $0.1$ to $0.6$~eV, and keeps its sign.

Tellurium is the canonical elemental chiral crystal: helical chains of atoms
wind around screw axes in space group $P3_121$ (or its enantiomer $P3_221$),
it shows a radial spin texture, current-induced bulk magnetization, and
chirality-dependent charge-to-spin
conversion.\cite{Sakano2020,Furukawa2017,Calavalle2022}  Our electronic
Hamiltonian is constructed from the vertex-corrected quasiparticle
self-consistent $GW$ (\QSGWhat{}) bands including spin--orbit
coupling.\cite{vanSchilfgaarde2006,Kotani2007,Pashov2020:CPC,Cunningham2023}  It differs from
standard $GW$ in that the potential is calculated self-consistently, and it also
carries the electron--hole ladder vertex in both the self-energy and the
screened interaction.  The combination of these extensions yields a
high-fidelity one-electron Hamiltonian, shown to be accurate for hundreds of
insulators.\cite{Runge2026:npj}  The computed gap is $\approx285$~meV.  The only reported gap~\cite{Shinno1973:JPSJ} is a magneto-optical absorption (excitonic) feature, which probes optical rather than quasiparticle excitations.

The electron--phonon interaction is calculated by the frozen-phonon approach,
which requires the change in the electronic Hamiltonian for a nuclear
displacement.  Generating the full \QSGWhat{} self-energy for each displaced
geometry is prohibitively expensive, so we use a variant of the Rigid
Self-Energy Approximation, which assumes that the self-energy shift relative
to DFT, $\Sigma(\bm{r},\bm{r}')-V^{\mathrm{DFT}}_{\mathrm{xc}}$, shifts rigidly
with the displacement of the atomic centres (analogous to the rigid-ion
approximation in electron--phonon theory), rather than being re-evaluated
self-consistently at every perturbed nuclear
position.\cite{Yin2011}  The density and potential are made
self-consistent for each displacement $\bm{u}$, so the main approximation is
that the shift in the effective potential is accurately carried by
$\Delta(V^{\mathrm{H}}+V^{\mathrm{DFT}}_{\mathrm{xc}})$ from DFT, equivalent
to approximating the \QSGWhat{} susceptibility by the DFT one for the purpose
of this shift.\cite{Acharya2021}  We map the electronic Hamiltonian to a
real-space, maximally localized Wannier basis of $2\times9$
orbitals;\cite{Marzari1997,Mostofi2014} the
projection is essentially exact, since Te $p$ orbitals are the only states
near $E_F$.  The diagonal part of the electron--phonon self-energy is kept,
obtained from the shift in the eigenvalue,
$\partial\varepsilon/\partial\bm{u}$.  This places the band ordering and the
spin texture on a first-principles
footing.  The electron--phonon vertices are
frozen-phonon deformation potentials $\partial\mathcal{H}/\partial u$ computed in the
same Wannier gauge, so the coupling and the bands are mutually consistent; the
on-site terms (the diagonal $\partial\varepsilon/\partial u$) are
$0.3$--$0.7$~eV/\AA{} in magnitude across the Wannier basis with the full $18\times18$ coupling matrix for every hopping
given in the Supplemental Material.
The phonon eigenvectors are taken from force constants on the identical geometry
along the $\Gamma$--A line of the screw axis.

Two features of the phonons are essential.  First,
at the zone centre $\Gamma$ every optical branch has vanishing phonon angular
momentum $l_z$ by point-group symmetry, so a $\Gamma$-point phonon gives an
identically null selection and contributes nothing; the genuinely chiral modes
live at finite $q_z$ along $\Gamma$--A, where the screw symmetry endows them
with a nonzero $l_z$.\cite{ZhangNiu2014} Second, the helicity partner of a
finite-$q_z$ mode is the mode at $-q_z$, obtained by the same $q\rightarrow-q$
operation as in the model.  We form a wire by stacking $N$ cells along the
screw axis at fixed transverse momentum $\bm{k}_\perp$ near the H point, where
the band edges lie.  The phonon wavevector points along $\Gamma$--A, so it conserves $\bm{k}_\perp$ and couples electron states at the same $\bm{k}_\perp$.  A bulk observable is a sum over $\bm{k}_\perp$, and we
sample it with 19 points, the H point and rings of 6 and 12 points at distances $0.03$ and $0.06$ (in units of the reciprocal-lattice vectors) from it; we
report the mean over these points.  A bias enters through
wide-band leads exactly as before, and the helicity dampings $\gamma_{\pm q}$
are again the Migdal bubble on the driven nonequilibrium Green's functions,
with the physical signal the bias-odd part $\dgamma_{\mathrm{odd}}$.

Figure~\ref{fig:abinitio} shows the two \textit{ab initio} ingredients this
construction uses directly: the
\QSGWhat{}$+$SOC band structure, colored by the spin expectation value
$\langle S_z\rangle$ [Fig.~\ref{fig:abinitio}(a)], from which the wire Hamiltonian and its
Wannier-gauge deformation potentials are built; and the chiral $\Gamma$--A
phonon dispersion, colored by the angular momentum $l_z$ of each branch's
eigenvector [Fig.~\ref{fig:abinitio}(b)], from which the traveling vertex $A_q$ is built.
The valence and conduction band edges sit within a few percent of the H
point, the pocket the wire is built near, and six of the nine
$\Gamma$--A branches carry $|l_z|\gtrsim0.3$ at $q_z=1/3$ (filled circles)
and are used below; the remaining three (crosses) do not. In the left-handed crystal every chiral branch has the opposite $l_z$, $l_z(\text{left})=-l_z(\text{right})$ to better than $10^{-8}$, as the mirror symmetry requires. The full
electronic structure, Wannier construction, frozen-phonon coupling formula,
and branch-by-branch $l_z$ table are given in the Supplemental Material.

\begin{figure*}[!tp]
\includegraphics[width=\textwidth]{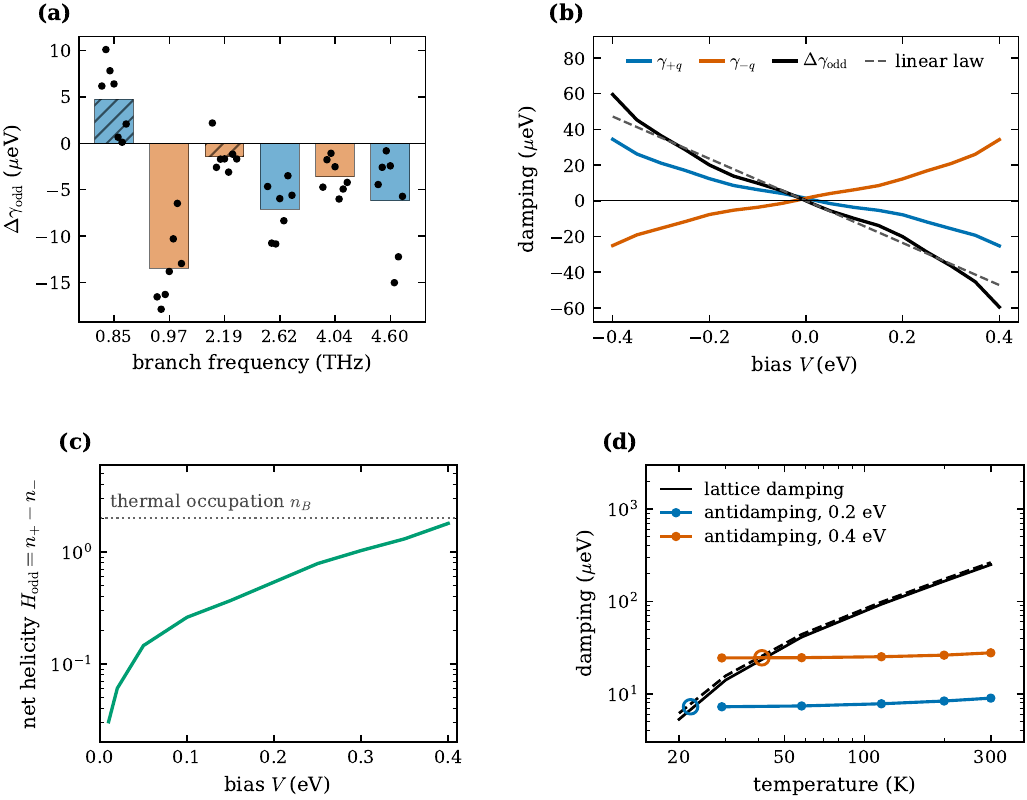}
\caption{Elemental tellurium from first-principles electronic structure and phonons.
(a)~Bias-odd damping asymmetry $\dgamma_{\mathrm{odd}}$ at $V=0.2$~eV for the six chiral $\Gamma$--A branches, labelled by frequency and colored by the sign of the phonon angular momentum $l_z$ (blue $l_z>0$, orange $l_z<0$); bars show the mean and dots the seven transverse momenta, $N=16$.  Hatched: the 0.85~THz branch depends on the wire length and the 2.19~THz branch changes sign across the transverse momenta.
(b)--(d)~Branch at 0.97~THz ($l_z=-0.94$), $N=16$, 116~K unless stated; mean over 19 transverse momenta.
(b)~Damping of the two helicities, $\gamma_{+q}$ and $\gamma_{-q}$, and the bias-odd asymmetry $\dgamma_{\mathrm{odd}}$ versus bias; the dashed line is the linear law fitted for $|V|\leq20$~meV.
(c)~Net helicity $H_{\mathrm{odd}}$ from the rate equation with the lattice damping from phono3py for two parameter sets of the machine-learned potential (band); the dotted line is the thermal occupation $n_B$.
(d)~Lattice damping versus temperature (solid and dashed: the two parameter sets) and the largest electronic antidamping at $V=0.2$ and $0.4$~eV; the coherent regime is reached where the antidamping exceeds the lattice damping, at the crossings marked by open circles (about 22 and 40~K).\label{fig:te}}
\end{figure*}

Figure~\ref{fig:te}(a) compares the six chiral $\Gamma$--A branches at $V=0.2$~eV.  On the branches at 0.97, 2.62, 4.04 and 4.60~THz the current selects the same helicity at every transverse momentum, with $\dgamma_{\mathrm{odd}}<0$ and a mean between $-3.6$ and $-13.5$~$\mu$eV.  The 0.85~THz branch has the opposite sign ($+4.8$~$\mu$eV) and depends strongly on the length of the wire.  On the 2.19~THz branch the mean is $-1.4$~$\mu$eV and the sign changes across the transverse momenta, from $-3.1$ to $+2.2$~$\mu$eV.  We analyse the 0.97~THz branch ($l_z=-0.94$, $\hbar\omega_0=4.0$~meV), which has the largest effect, $-13.5$~$\mu$eV, and is the best converged.

For this branch the dampings $\gamma_{+q}$ and $\gamma_{-q}$ are equal at zero bias and split with the current [Fig.~\ref{fig:te}(b)].  The odd asymmetry is linear in the bias at small bias, $\dgamma_{\mathrm{odd}}=\chi V$ with $\chi=-1.2\times10^{-4}$ (a pure number, $V$ in volts, $\dgamma$ in eV), and $\chi$ is negative at every transverse momentum [Fig.~\ref{fig:te}(b)].  At $V=0.2$~eV, $\dgamma_{\mathrm{odd}}$ at the H point is between $-2.1\times10^{-5}$ and $-1.4\times10^{-5}$~eV for wires of 12 to 32 cells.  The slope changes from $-1.3\times10^{-4}$ at 30~K to $-1.0\times10^{-4}$ at 300~K.  The asymmetry of about $2\times10^{-5}$~eV at $0.2$~V follows from the computed deformation potentials of $0.3$--$0.7$~eV/\AA{}.

The kinetics of Eq.~\eqref{eq:rate} needs the lattice damping $\gamma_{\mathrm{lat}}$.  We compute it with phono3py from the third-order force constants of the same machine-learned potential, for this branch at the same wavevector $q_z=1/3$.  It is $93$--$99$~$\mu$eV at 116~K and $250$--$264$~$\mu$eV at 300~K for two parameter sets of the potential that agree within about 10\% from 30 to 300~K.  The same calculation gives the zone-centre optical modes of tellurium within about 15\% of the measured Raman frequencies and with linewidths of the size measured.\cite{Spirito2024:JMCC}  With this lattice damping the current produces a net helicity $H_{\mathrm{odd}}=0.06$ at $20$~mV and $0.5$ at $0.2$~V, $3\%$ and $26\%$ of the thermal occupation $n_B=2.0$ of the mode [Fig.~\ref{fig:te}(c)].

The selection persists to room temperature: at $0.2$~V the net helicity is $0.5$--$0.7$ for every temperature from 58 to 300~K.  The coherent regime appears when the electronic antidamping exceeds the lattice damping [Fig.~\ref{fig:te}(d)].  Averaged over the transverse momenta, the antidamping of the favoured helicity is $7$--$9$~$\mu$eV at $0.2$~V and $25$--$28$~$\mu$eV at $0.4$~V, nearly independent of temperature, whereas $\gamma_{\mathrm{lat}}$ falls with temperature.  The two cross at about $22$~K for $0.2$~V and about $40$~K for $0.4$~V, which is $4\times10^{7}$~V/m across the 16-cell wire.  Above these temperatures the lattice damping exceeds the electronic antidamping at every bias computed, and the helicity takes the finite value given by Eq.~\eqref{eq:nss}.





\section{Discussion}\label{sec:discussion}
We have shown that a biased chiral conductor can populate a chiral phonon without
optical pumping, a magnetic field, or a tuned resonance: the charge current
selects the helicity.  Helicity selection requires structural chirality together
with finite-$q$ lattice modes that carry a travelling vertex
$A_q\neq A_{-q}$.  Ordinary conductors lack that combination: inversion, mirror,
or glide symmetries that map $q\rightarrow-q$ force $\dgamma\equiv0$ at any
bias, which is why the achiral Rashba wire of Sec.~\ref{sec:results} carries a
current with zero helicity selection, and zone-centre modes have $l_z=0$ by
point-group symmetry and give a null.  A nonequilibrium steady state is
shared by every current-carrying conductor, but which of those conductors convert
current into a definite phonon helicity is fixed by Neumann's principle for a
time-reversal-odd response tensor.  Chirality lifts that symmetry
restriction.~\cite{Jacquod2012:PRB}
Onsager reciprocity constrains linear response about thermal
equilibrium~\cite{Onsager1931} and is fully respected here: at $V=0$,
time-reversal symmetry enforces $\gamma_{+q}=\gamma_{-q}$, so $\dgamma(0)=0$
and $H=0$.  A finite bias is itself time-reversal odd, and the response must
reverse under $V\rightarrow-V$, which Eq.~\eqref{eq:Dgamma} encodes
($\dgamma$ is the bias-antisymmetric part).  The leading term
$H_{\mathrm{odd}}\propto V$ as $V\rightarrow0$ is the first term of that odd
expansion.  Above $V_c$ the soft helicity runs away into a coherent amplitude;
Eq.~\eqref{eq:nss} diverges smoothly as $V\rightarrow V_c^-$, and
$\dgamma(V)$ itself stays continuous through the threshold.  A phonon-mediated
CISS polarization is parallel, while an ordinary electronic
Edelstein accumulation is conventionally transverse.  (Materials with
lower-symmetry spin--orbit coupling can support a longitudinal electronic
channel as well, so geometry is a useful guide rather than a sharp separator.)

The subthreshold helicity is linear and odd in the current, so sweeping $V$
through zero sends $H_{\mathrm{odd}}$ continuously through the origin and
reverses its sign; at $V=0$ the system is back in equilibrium with no leftover
phonon helicity.  What sustains the nonequilibrium state at finite $V$ is the
open boundary condition: the leads sit at different chemical potentials, a
steady current flows, and energy is exchanged with the reservoirs, while the
Hamiltonian remains Hermitian and time-reversal symmetric.  Turn the bias off
and both the electronic distribution and the phonon helicity relax on ordinary
dissipative timescales.  Every propagator entering $\dgamma$ is the
stationary, time-translation-invariant solution of the driven NEGF equations
in the energy domain, so the sign, size, and threshold of the selection are
read directly from that steady state.  The physical signal is the odd part of
$H$; small even residuals in finite open wires or from approximate
electron--phonon vertices are projection artifacts.

Below $V_c$, the lattice helicity is linear in bias at small $V$
($H_{\mathrm{odd}}\propto V$ from Eq.~\eqref{eq:nss}).  Fitting
$\dgamma(V)=aV+cV^3$ over the subthreshold window gives a clear negative cubic
coefficient: $|cV^3|$ becomes comparable to $|aV|$ well before $V_c$, so the
damping asymmetry softens with bias while remaining odd in $V$ by time
reversal.  That same odd-in-current structure is what most CISS
magnetoresistance measurements report at modest
bias.\cite{Bloom2024,Huisman2023}  (Higher harmonics of a measured
magnetocurrent can also track the lead
model.\cite{Huisman2023:JCP})  Only the coherent regime has a threshold.

Three signatures follow.  Reversing the current reverses the selected
helicity and any phonon-linked spin signal.  Reversing the enantiomer
reverses the axial spin that spin--orbit coupling attaches to the selected
phonon, while the laboratory helicity at fixed $q$ stays put, so the CISS
polarization flips with handedness.  An achiral conductor shows
no selection at any bias.  These map directly onto Fig.~\ref{fig:schematic}.

Identifying $\dgamma$ with the current-induced Berry force connects the
effect to nonconservative forces and current-driven vibrational
instabilities in nanoscale
conductors;\cite{Dundas2009,Todorov2010,Lu2010,Bode2011,Lu2012} in a chiral
conductor the coupled coordinates are the two circular polarizations of a
phonon, so the force selects helicity, and its strong-drive limit is a
coherent chiral phonon.  The same reading connects our result to the
phonon Edelstein effect\cite{Yokoyama2025,Yokoyama2026} and the phonon rotoelectric
effect,\cite{Hamada2018,Hamada2020} which describe current- and
temperature-driven phonon angular momentum; the subthreshold regime here is
the transport member of that family, evaluated on a driven electronic
steady state.  Converting the selected helicity into a spin polarization or magnetoresistance requires one further step, the imprint of the selected phonon on the driven electrons, which is treated for the model in Ref.~\cite{AcharyaUnified2026}.

The phonon Einstein--de Haas effect is related but distinct.  The standard
definition of phonon angular momentum,
$l_\sigma(k)=\hbar\,\epsilon_\sigma^{\dagger}(k)M\epsilon_\sigma(k)$ with $M$ the
generator of rotations about the chiral axis,\cite{ZhangNiu2014} is exactly the
on-site circular ionic motion of mode $(k,\sigma)$; time reversal enforces
$l_\sigma(k)=-l_\sigma(-k)$, so this quantity, like $\dgamma$ in this work, vanishes at
equilibrium.  A temperature gradient unbalances it in nonmagnetic chiral
crystals,\cite{Hamada2018} and this phonon Einstein--de Haas effect has now
been measured directly in single-crystal tellurium as a mechanical torque that
reverses with gradient direction and enantiomer.\cite{Miao2025}  The tellurium
branches highlighted in our results have large $|l_z|$ because they are the
material's chiral modes at finite $q_z$.  Related work computes the
mechanical torque that follows from the current-induced selection established
here in tellurium's chiral modes, the electrical analog of the measured
thermal Einstein--de Haas torque.~\cite{AcharyaEDH2026}

\section{Conclusions}\label{sec:conclusions}
In this paper we establish that a dc electric current selects the helicity of a
chiral phonon in a chiral conductor.  The bias-antisymmetric damping $\dgamma$ is
the nonconservative Berry force on the phonon coordinate; with a phonon rate
equation it produces a net lattice helicity at any finite bias and, above a
threshold, a self-sustained coherent mode of definite $q$.  That closes the
missing step for the coherent-phonon route to CISS that we identified in Ref.~\cite{AcharyaUnified2026}:
in that work a coherent mode was taken as an assumption; here it is a
byproduct of the theory.  A quasi-\textit{ab initio} study of elemental tellurium
with \QSGWhat{} for the electronic structure shows the
same selection for finite-$q$ chiral modes.  In tellurium the selection persists from 30~K to room temperature, and the coherent regime appears below about 40~K at the largest bias computed ($0.4$~V across a 9.5~nm wire).  This work determines, for a given
band filling and branch, which helicity is
populated, and the sign of the selection follows the current.  Reversing the bias reverses the
selected helicity and any phonon-linked spin signal.

\section{Methods}\label{sec:methods}
\subsection*{Helical-wire model}
Figure~\ref{fig:model} uses a helical wire of
$N=12$ sites (validation runs use $N=8$) with unit nearest-neighbor hopping
$t=1$ (the energy unit), chiral spin--orbit strength
$\lambda_{\mathrm{soc}}=0.4$, and a chiral phonon of wavevector $q=2\pi/5$ and
frequency $\omega_0=0.15$.  The electron--phonon vertex keeps the
hopping-modulation ($t_1=1$) and chiral spin--orbit-modulation
($\lambda_1=0.6$) channels active and switches the Holstein (on-site) channel
off ($\epsilon_1=0$).  The two active channels are comparable in magnitude and
interfere with each other, so both are needed to evaluate $A_q$ correctly.  The
electronic part is solved in a wide-band nonequilibrium Green's
function scheme\cite{MeirWingreen1992,JauhoWingreenMeir1994,HaugJauho2008} at
temperature $k_BT=0.02$.  The lead self-energies $\Gamma_{L,R}$ are fixed,
bias-independent broadenings; bias enters only through the two reservoirs'
Fermi functions, so at $V=0$ ($\mu_L=\mu_R$) they coincide and the
Meir--Wingreen current is exactly antisymmetric under exchanging
$L\leftrightarrow R$, vanishing identically rather than by cancellation of a
small residual. We confirm $I(V{=}0)=0$ to machine precision in every
calculation reported here, despite the leads being the only non-Hermitian
part of the problem.  The phonon self-energies of Eq.~\eqref{eq:gamma} are
evaluated on the driven $G^{\gtrless}(E)$ with the energy grid locked to
$\omega_0$ so that the shift $E\pm\omega_0$ is exact on the
grid.\cite{Frederiksen2007}  For these parameters the equilibrium damping is
$\gamma^e(0)\approx0.10$, the soft helicity reaches antidamping with
$\Delta_{\max}\approx1.5\times10^{-2}$ near $V\approx1.35$, and the electronic
instability sits at $V_c\approx1.24$.  The rate equation~\eqref{eq:rate} uses a
representative lattice damping $\gamma_{\mathrm{lat}}=5\times10^{-3}$ for
Fig.~\ref{fig:model}(c); the phase boundary of Fig.~\ref{fig:model}(d) sweeps
$\gamma_{\mathrm{lat}}$ across $\Delta_{\max}$.  The Berry force of
Eq.~\eqref{eq:identity} is computed independently as the antisymmetric part of
the electronic friction matrix in the standing-wave basis and agrees with
$\dgamma$ to the full working precision at every bias.

\subsection*{Tellurium}
The electronic structure is a maximally-localized Wannier
interpolation of vertex-corrected quasiparticle self-consistent $GW$
(\QSGWhat{}) bands with spin--orbit
coupling.\cite{vanSchilfgaarde2006,Kotani2007,Cunningham2023,Marzari1997,Mostofi2014}
Electron--phonon vertices are frozen-phonon deformation potentials
$\partial\mathcal{H}/\partial u$ in the same gauge.  Phonon eigenvectors and frequencies
are obtained from force constants on the identical geometry; the chiral modes
are taken at finite $q_z$ along $\Gamma$--A, and their angular momenta $l_z$ are
evaluated from the eigenvectors.  The full electronic structure, the Wannier
construction and its gauge-consistency check, the frozen-phonon coupling
formula, and the frequency and $l_z$ of every $\Gamma$--A branch are given in
the Supplemental Material.  The transport wire stacks $N$ cells along the screw axis ($N=16$ for the results shown, $N=12$ to $32$ in the length tests).  The transverse momentum is sampled at 19 points, the H point and rings of 6 and 12 points at distances $0.03$ and $0.06$ from it (in units of the reciprocal-lattice vectors); the small-bias slope of the asymmetry changes by 2\% between these 19 points and the 7 points of the first two shells, which are used for the branch comparison.  The chemical potential is set $50$~meV below the bulk valence-band maximum at each transverse momentum, wide-band leads with coupling $\gamma_0=0.3$~eV carry the bias, and $k_BT=0.01$~eV ($116$~K) unless stated.  A phonon is damped by all electrons, so $\gamma_{\pm q}$ and $\Gamma^{\mathrm{em}}_\pm$ are averaged over the transverse momenta before the rate equation is closed.  The lead coupling changes the asymmetry by a factor of 2.6 between $\gamma_0=0.1$ and $0.6$~eV with the same sign ($\dgamma_{\mathrm{odd}}=-8.0,\,-16.5,\,-21.0\times10^{-6}$~eV for $\gamma_0=0.1,\,0.3,\,0.6$~eV at the H point, $V=0.2$~eV).  The lattice damping is the phonon decay rate from phono3py with the machine-learned potential ($3\times3\times3$ supercell, displacement $0.03$~\AA{}, pair cutoff $6$~\AA{}, $12\times12\times12$ $q$ mesh) at $q=(0,0,1/3)$, twice the half-width that phono3py reports, for 20, 30, 58, 116, 200 and 300~K.  A $18\times18\times18$ mesh, an $8$~\AA{} cutoff or a $4\times4\times4$ supercell changes it by up to about one third for the two parameter sets used (\texttt{omat\_pbe} and \texttt{mp\_pbe\_refit\_add}).  The inputs beyond first principles are the transport parameters stated here.

\section*{Data Availability}
The electronic-structure and phonon data reported here were produced with
standard open-source packages: Questaal,\cite{Pashov2020:CPC} janus-core,\cite{JanusCore2024} phonopy and phono3py.\cite{Togo2015:Phonopy,Togo2023} All input and output files will
be provided by the corresponding author upon reasonable request.

\section*{Author Contributions}
S.A.\ conceived the study, developed the GKBA and SCBA codes and the phono3py interface with janus-core,
performed the \QSGWhat{} calculations and all other calculations, and wrote the paper.  M.v.S.\ and J.J.\ contributed software and
the writing of the draft.

\begin{acknowledgments}
This work was authored by the National Laboratory of the Rockies for the U.S.
Department of Energy (DOE) under Contract No.\ DE-AC36-08GO28308.  Funding for S.A.\ and
M.v.S.\ was provided by the Computational Chemical Sciences program
within the Office of Basic Energy Sciences, U.S.\ Department of Energy.  SA acknowledges the use of the National Energy
Research Scientific Computing Center, under Contract No.\ DE-AC02-05CH11231 using
NERSC award BES-ERCAP0021783, and also acknowledges that a portion of the research
was performed using computational resources sponsored by the Department of
Energy's Office of Energy Efficiency and Renewable Energy and located at the
National Laboratory of the Rockies.  The views expressed in the article do not
necessarily represent the views of the DOE or the U.S. Government.  The U.S.
Government retains and the publisher, by accepting the article for publication,
acknowledges that the U.S. Government retains a nonexclusive, paid-up,
irrevocable, worldwide license to publish or reproduce the published form of this
work, or allow others to do so, for U.S. Government purposes.
\end{acknowledgments}

\bibliography{references}

\end{document}